\documentclass[journal=jacsat,manuscript=article]{achemso}
\usepackage{amsfonts}
\usepackage{optidef}
\usepackage{lipsum}  
\usepackage[version=3]{mhchem} % Formula subscripts using \ce{}
\usepackage{lineno}
\title{Multi-dimensional wavefront sensing using volumetric meta-optics}

\author{Conner Ballew, Gregory Roberts, and Andrei Faraon}

\affiliation{Kavli Nanoscience Institute and Thomas J. Watson Sr. Laboratory of Applied Physics, California Institute of Technology, Pasadena, California 91125, USA}

\email{faraon@caltech.edu} %% email address is required

\begin{document}

%\title{Volumetric metaoptics for multi-functional applications}

%\title{Volumetric meta-optics for efficient mapping of direction wavelength and polarization on a 2D image sensor}



\begin{abstract}
The ideal imaging system would efficiently capture information about all fundamental properties light: intensity, direction, wavelength, and polarization. Most common imaging systems only map the spatial degrees of freedom of light onto a two dimensional image sensor, with some wavelength and/or polarization discrimination added at the expense of efficiency. Thus, one of the most intriguing problems in optics is how to group and classify multiple degrees of freedom and map them on the two dimensional sensor space. Here we demonstrate through simulation that volumetric meta-optics elements consisting of a highly scattering, inverse-designed medium structured with subwavelength resolution can sort light simultaneously based on direction, wavelength and polarization. This is done by mapping these properties to a distinct combination of pixels on the image sensor for compressed sensing applications, including wavefront sensing, beam profiling, and next-generation plenoptic sensors.
\end{abstract}

%%%%%%%%%%%%%%%%%%%%%%%%%%  body  %%%%%%%%%%%%%%%%%%%%%%%%%%
\section{Introduction}

Two dimensional image sensors are the most common detectors for light, so a leading optical engineering task is how to best extract the information from the incident optical field using an optical system and a planar image sensor. For example, a black and white camera maximizes the amount of spatial information by performing a one to one mapping between the direction of propagation and a pixel on the image sensor, but all other information is lost. A line scan multispectral camera maps only one spatial coordinate to a direction on the image sensor while the other direction records the spectrum of the spatial pixel. Various other mappings are used in cameras to image color, polarization, light field, etc. Since the information capacity of a 2D image sensor is finite, getting more information about some degrees of freedom for light comes at the expense of information in other degrees of freedom. Also, in general purpose systems with trivial mapping implementations that use conventional optical components like lenses, gratings, and prisms, a single pixel on the image sensor detects a specific combination of single degrees of freedom.  For example, a camera may detect the combination of a specific direction, wavelength band, and polarization. Thus, if $S$ spatial directions, $W$ wavelength bands, and $P$ polarizations need to be resolved, then $S \times W \times P$ pixels are needed. However, many times there is prior knowledge about the input light field, in which case it is possible to use mappings that more efficiently utilize the pixels of the sensor \cite{Suzen2010,Pegard2016,Mait2018,https://doi.org/10.48550/arxiv.1905.13221, Roberts2022}. Here, we demonstrate an inverse designed volumetric meta-optic device that can efficiently map different combinations of wavelengths, directions and polarizations into combinations of pixels on an image sensor. The compressed information can fully classify properties of the incident fields under certain approximations, namely that the wavefronts are monochromatic, locally linear in phase, and linearly polarized.

The device is based on metaoptics, which describes materials patterned with subwavelength resolution that impart customized transformations to incident light. Most research in meta-optics focuses on dielectric metasurfaces that consist of a single, approximately wavelength-thick layer of subwavelength scale antennas with the ability to completely control the phase and polarization of an incident wavefront \cite{Yu2014,Arbabi2015,Chen2016}. The ability to impose independent phase profiles to two orthogonally polarized inputs allows a single metasurface to perform tasks that historically could only be achieved with cascaded systems of bulk components, which has improved applications that must obey strict size constraints while maintaining high efficiencies. However, further expanding the multifunctionality of metasurfaces to multiple angles and wavelengths tends to come at the cost of efficiency \cite{Miller2007, Arbabi2017}. 

To recover the performance lost by increasing multifunctionality, the size of the metaoptics system can be increased. This idea has been explored in recent years by employing multiple metasurfaces in a cascaded system \cite{Pfeiffer2013, Avayu2017, Zhou2018, Huang2019, Mansouree2020}. While these systems have a reduced size relative to systems that employ only conventional bulk optical elements, the size of the system is primarily dictated by the distance between metasurfaces. The distance between metasurface elements must be sufficiently large to preserve the assumptions that metasurface design is reliant on, which are described in detail in Ref. \cite{Kamali2018, Miller2023}. The paradigm of cascading elements that are spatially separated enough to preserve their design independence provides an intuitive way to design complex systems, but is not strictly necessary. Instead, design methodologies that account for the effects of near-field coupling and multiple scattering events (i.e. full-wave Maxwell solvers, including FDTD) are required for more advanced operation.

Recent work has demonstrated the tractability of designing 3D volumetric metaoptics using inverse-design techniques. These techniques are aided by the adjoint method for electromagnetics, which utilizes adjoint symmetry in Maxwell’s equations to efficiently compute the gradient of arbitrary figures of merit (FoMs) with respect to material permittivity \cite{Miller2012,Lalau-Keraly2013,Molesky2018}.  This can be subsequently used to optimize the shape of a structure in a process referred to as topology optimization \cite{Jensen2004,Jensen2011}. Full-wave simulations require significant computational resources to perform, so applications of inverse-designed metaoptics have been limited to integrated waveguide components \cite{Piggott2015} or small 3D components \cite{Camayd-Munoz2020}, on the order of several wavelengths per side.

For this work, we investigate through simulation a theoretical extreme for the metaoptics platform by designing a device that can classify incident light based on all of its fundamental properties: intensity, propagation direction, wavelength, and polarization. To our knowledge this amount of multi-functionality at both high efficiency and miniaturization has neither been obtained nor investigated to this degree in the past. The intent of this study is to explore the possibilities ahead for volumetric metaoptics, provide a meaningful goal for future photonic foundry processes, and pose a device that could be of high interest to the computational imaging community.

Due to the computational difficulty of this task, we target a platform in which a periodic array of the designed 3D devices can be placed above a sensor array. This allows the devices to be small during the design, then tiled to cover a desired area. The devices are designed for three functionalities: polarization splitting of two linearly polarized states, wavelength splitting of two distinct wavelengths, and a basic form of imaging in which five planewave components are focused to five different locations. This totals to nine states we seek to classify, so in practice the device can be placed above a 3×3 grid of pixels. Although the device is designed for a discrete set of states, it can be used to interpolate states on a continuum by analyzing the ratio of intensities among all the pixels.

\begin{figure}[htbp]
\centering\includegraphics[width=12cm]{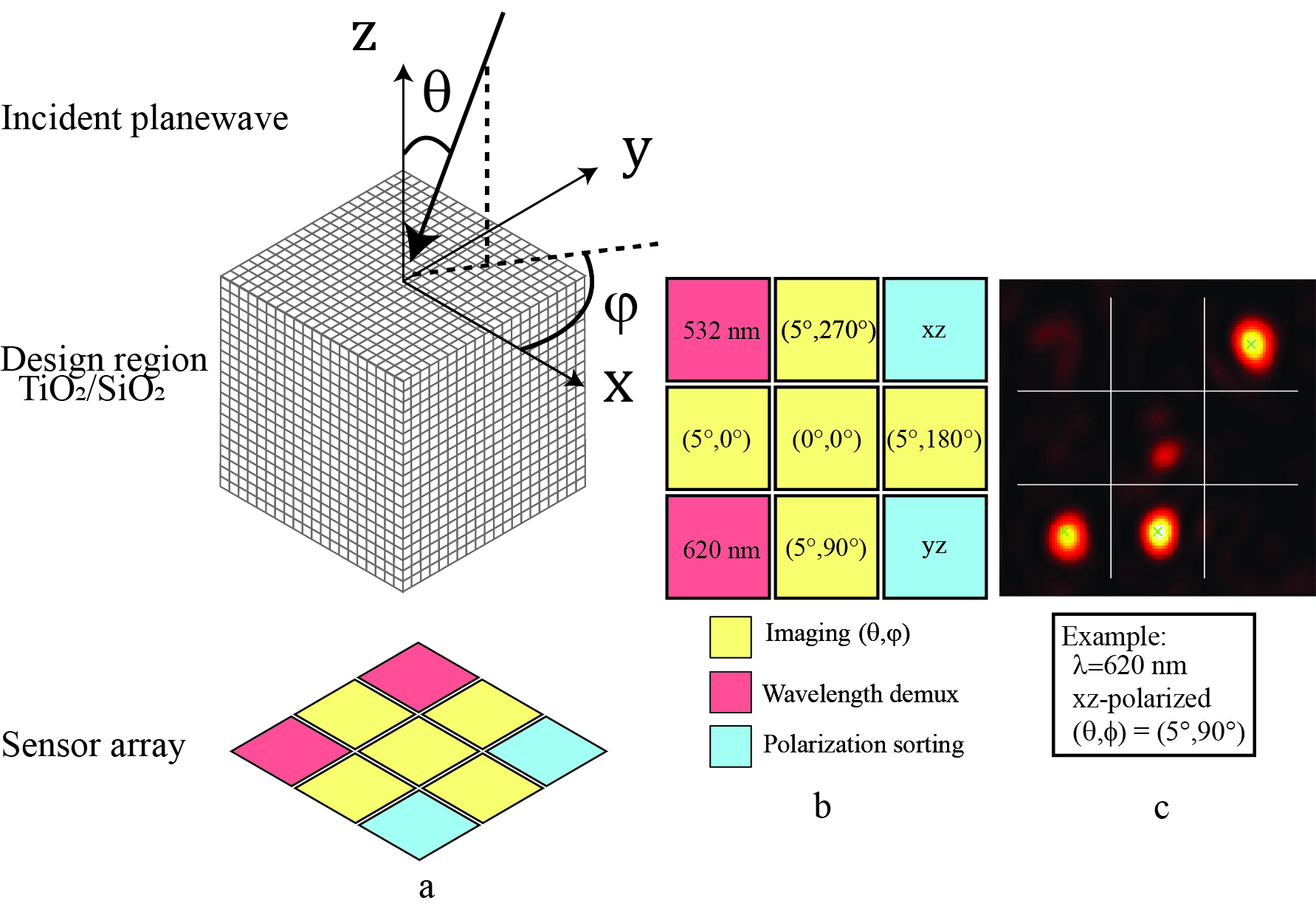}
\caption{\textbf{The layout of the system, which features a device region placed above an array of sensors.} (a) An incident planewave is input to a device comprised of SiO$_2$ and TiO$_2$ at an angle $(\theta, \phi)$. The bottom of the device is 1.5 $\mu m$ above a sensor array. The background material is assumed to be SiO$_2$. (b) The distribution of functionalities across the pixel array. The device focuses light to different pixels depending on the state of the input light. (c) The output of 620 nm, xz-polarized planewave input incident at an angle $(\theta,\phi)=(5^{\circ},90^{\circ})$ }
\label{Fig1}
\end{figure}

The polarization and wavelength splitting functionalities are designed to work for all of the assumed incident angles, which include a normally incident case and non-normal incidences that are 5 degrees tilted from normal (polar angle $\theta=5^{\circ}$) with azimuthal angles $\phi\in\{0^{\circ},90^{\circ},180^{\circ},270^{\circ}\}$. The two design wavelengths are 532 nm and 620 nm. To prevent the structure from being highly resonant, frequencies within a $\pm 5$ THz range (approximately $\pm 4$ nm) of these design frequencies are also optimized. The two polarizations are linear and orthogonal, with one polarized in the xz-plane and the other in the yz-plane. The device is a 3µm × 3µm × 4µm stack of 20 layers. Each 200 nm layer is comprised of titanium dioxide (TiO$_2$, n=2.4) and silicon dioxide (SiO$_2$, n=1.5), with a minimum feature size of 50 nm. The background material is assumed to be SiO$_2$. A schematic of the device, sensor array, and examples of the fields at the focal plane under a 620 nm xz-polarized planewave excitation angled at $(\theta,\phi) = (5^\circ,90^\circ)$ is shown in Fig. \ref{Fig1}. Note that since this $5^\circ$ polar angle is in SiO$_2$ with $n=1.5$, these planewave components couple to $7.5^\circ$ planewave components in air.

We begin by describing the inverse design process used to optimize the devices. Next, the performance of the device under the assumed input conditions is studied. Following this, we observe that the behaviors of functionalities are preserved and predictable when excited at states in between the states the device was optimized for. For example, as the input wavelength is continuously shifted from 532 nm to 620 nm, the ratio of the 620 nm pixel transmission to the 532 nm pixel transmission monotonically increases, while the imaging and polarization functions remain efficient. This occurs despite the device only being optimized in a narrow wavelength range around 532 nm and 620 nm. Similar behavior is observed in the imaging functionality when excited at input angles that were not explicitly optimized within a $5^{\circ}$ cone. The read-out from these pixels can thus be used to infer the state of the light incident on the device so long as the input is assumed to be a monochromatic planewave. We find that this behavior does not depend on the chosen mapping of functionality to specific pixels in the focal plane by analyzing the behavior of devices optimized for 20 unique pixel distributions.

\section{Methods}

The goal of photonic topology optimization is to find a refractive index distribution that maximizes an electromagnetic figure-of-merit (FoM). Since the device presented here is highly multi-functional, its optimization is multi-objective. Each objective is a mapping of an input to a FoM: the input is a planewave with a specific angle, polarization, and wavelength; the FoM is the power transmission through the desired pixel. The general procedure for this optimization is shown in Fig. S8 of the Supporting Information. It consists of three main steps: first, the FoMs and their associated gradients are computed \cite{Lalau-Keraly2013}; second, the gradients are combined with a weighted average \cite{Ballew2021}; third, the device is updated in accordance with the averaged gradient using either a density-based optimization of a continuous permittivity or a level-set optimization of a discrete permittivity \cite{Vercruysse2019}. 

The individual FoM gradients are evaluated at every point in the design region using the adjoint method, which entails combining the electric fields in the design region for a “forward” and an “adjoint” simulation to compute the desired gradient. In this case, the forward case simulates the device under the assumed planewave excitation, and the adjoint case simulates a dipole (with a particular phase and amplitude based on the forward simulation) placed at the center of the desired pixel. This choice of sources optimizes the device to focus light to the location of the dipole. However, we record the performance of the device as power transmission through the desired pixel rather than intensity at a point, since power transmission better represents the signal a sensor pixel would record. The remainder of this section elaborates on this procedure.

All unique forward and adjoint simulations are first simulated in Lumerical FDTD in parallel on a high-performance cluster. Next, the associated FoMs are recorded from the forward simulation results, and the FoM gradients are computed by combining the results of appropriate forward and adjoint source pairs. The individual gradients are spatially averaged in the z-direction for each layer, yielding a 2D gradient for each layer of the device. All FoM gradients are combined using a weighted average into a single gradient, still evaluated at every point in the design region. Information on this weighting procedure is described in detail in Ref. \cite{Ballew2021}. Finally, this interpolated gradient is used to update the permittivity of the device structure. 

The optimization is done in two phases: a density-based phase, and a level-set phase. Each phase has a unique update procedure. In the density-based optimization the permittivity of the device is modelled as a grid of grayscale permittivity values between the permittivity of the two material boundaries. This permittivity representation is effectively fictitious (unless an effective index material can be reliably fabricated), and the goal is to converge to a binary device that performs well and is comprised of only two materials. We use the methods described in Ref. \cite{Ballew2022} to achieve this.

While the density-based optimization can converge to a fully binary solution, it is faster to terminate the optimization early and force each device voxel to its nearest material boundary. This thresholding step reduces the device performance, which we recover by further optimizing the device with a level-set optimization. Level-set optimization models the device boundaries as the zero-level contour of a level-set function ($\phi(x,y) = 0$), and thus benefits from describing inherently binary structures \cite{Lebbe2019}. Empirically the final device performance is dependent on initial seed, hence the need for the improved density-based optimization that converges to a near-binary solution. Here we use level-set techniques to simultaneously optimize device performance and ensure the final device obeys fabrication constraints. This technique is described in detail in Ref. \cite{Vercruysse2019}, and elaborated on in the Supporting Information Section S2. The full convergence plot for the optimization and the resulting device layers are shown in Fig. S6 and Fig. S7, respectively, of the Supporting Information.

\section{Results}

We quantify the performance of the device in two ways. First, we check the performance of the device under the excitation beams that were assumed when optimizing the device, which we refer to as \textit{training modes}. Second, we study the device at different input angles, wavelengths, and polarization states to analyze its ability to classify states. We refer to these as \textit{validation modes}.

\begin{figure}[htbp]
\centering\includegraphics[width=14cm]{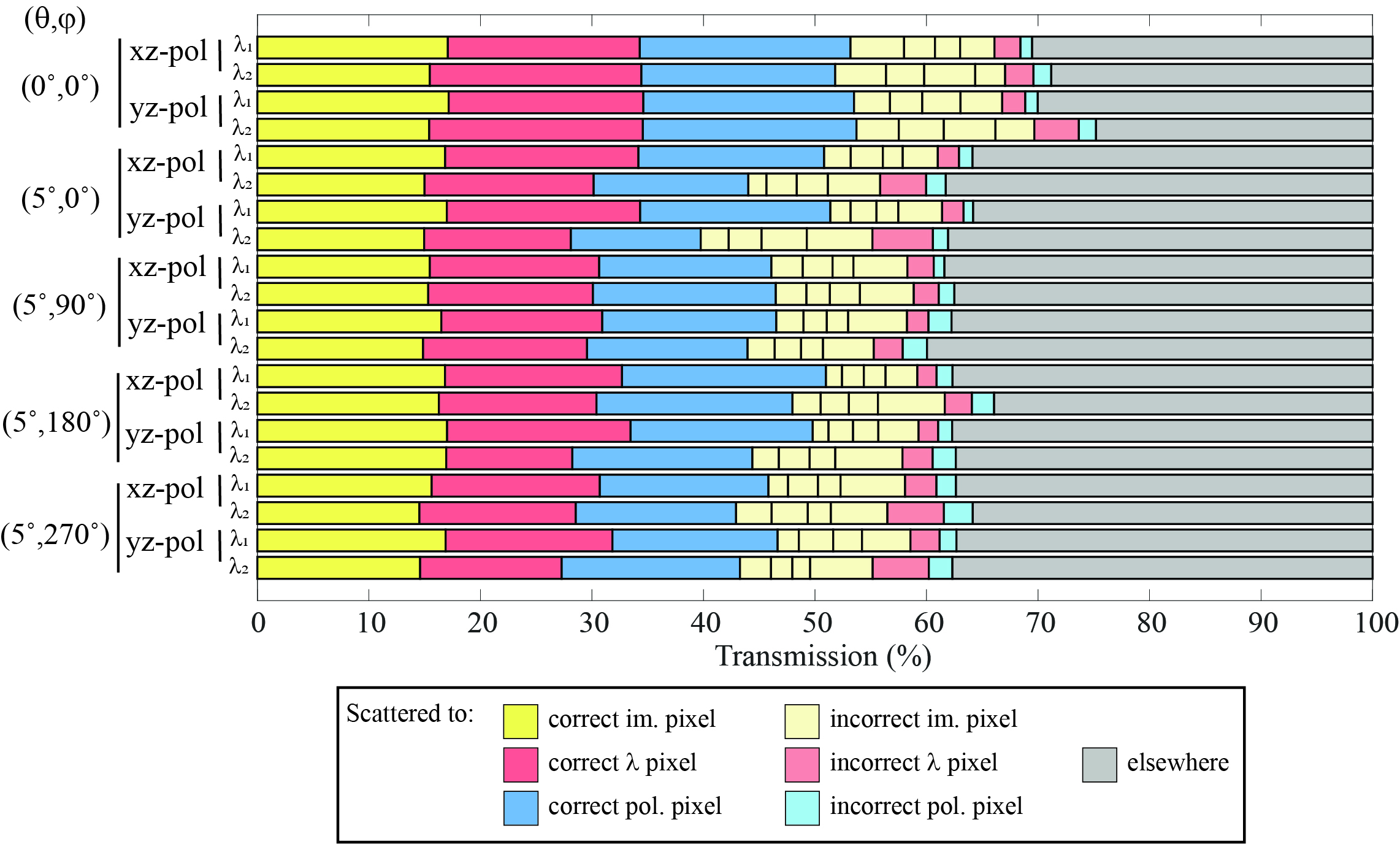}
\caption{The performance of the device under excitation of training modes. The input angle, polarization state, and wavelength are represented by the y-axis. The x-axis represents the fraction of power scattered to different regions. The bold colored bars quantify transmission to the correct pixel, while the pale colored bars quantify transmission to the incorrect pixel. The grey bar quantifies the power that is back-scattered or obliquely scattered, thus not reaching any of the pixels in the focal plane.}
\label{Fig3}
\end{figure}

The design methodology co-optimizes 20 different input excitations featuring unique combinations of wavelength, polarization, and angle of incidence. The transmission to each pixel for each input state is shown in Fig. \ref{Fig3}. For each state the transmission to the three correct pixels, the transmission to the six incorrect pixels, and the transmission elsewhere (oblique scattering, back-scattering, etc.) is shown. On average, the transmission to the correct pixels totals $47.7\%$, the transmission to incorrect pixels totals $16.8\%$, and the transmission elsewhere totals $35.5\%$.

The design process optimizes the performance of the device only for the training modes, and it is unclear what the output of the device will be if the excitation parameters are continuously varied. Here we quantify the performance of the device at input states between that of the designed input states. We study the performance at wavelengths in between 532 nm and 620 nm, at a large set of angles within a $10^\circ$ input angle cone, and at arbitrary polarization states.

\begin{figure}[htbp]
\centering\includegraphics[width=12cm]{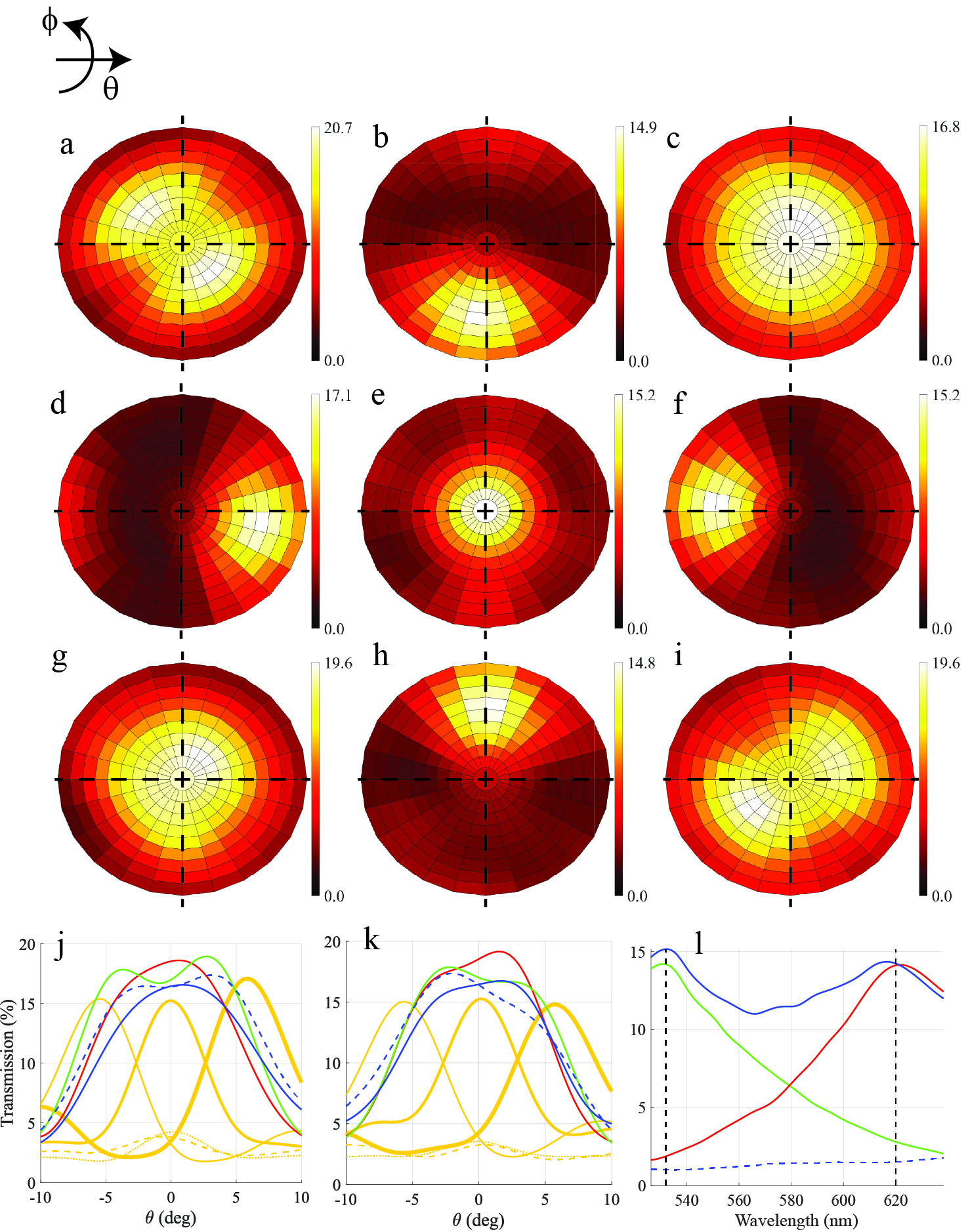}
\caption{The angular sensitivity of the device. (a)-(i) plot the transmission to the different pixels in the focal plane for $\theta$ from $0^\circ$ to $10 ^\circ$ and $\phi$ from $0^\circ$ to $360^\circ$. The top-down left-right ordering of these plots are matched to the ordering of the pixels depicted in Fig. \ref{Fig1}b. The wavelength demultiplexing plots (a) and (g) are averaged in polarization. The polarization sorting plots (c) and (i) are averaged in wavelength. The imaging plots (b), (d-f), and (h) are averaged in both polarization and wavelength. (j,k) are horizontal (j) and vertical (k) traces of the surface plots (a)-(i) along the black dashed lines. For (j) and (k), the line colors represent transmission to different pixels: the green line is the 532 nm pixel, the red line is the 620 nm pixel, the solid blue is the xz-polarization pixel, the dashed blue is the yz-polarization pixel, the solid yellow lines are the on-axis imaging pixels, and the dashed yellow lines are the off-axis imaging pixels. For (l), the green and red lines represent the transmission to the 532 and 620 nm pixels respectively. The solid blue line is the transmission to the correct polarization pixel given the polarization of the input source, and the dashed blue line is the transmission to the incorrect polarization pixel. For (l), all traces are averaged over all evaluated input angles within a $5^\circ$ cone.}
\label{Fig4}
\end{figure}

The full angular response of each individual pixel in a $10^{\circ}$ cone is shown in Fig. \ref{Fig4}a-i. The results along the horizontal dashed lines ($\phi=0^\circ$) and vertical dashed lines ($\phi=90^\circ$) are plotted in Fig. \ref{Fig4}j,k. To obtain these plots, the transmission values are averaged across wavelength, polarization, or both depending on the functionality. For the wavelength sorting functionality the values are averaged across polarization; for the polarization sorting functionality the values are averaged across wavelength; and for the imaging functionality the values are averaged across both wavelength and polarization. The wavelength and polarization demultiplexing functionalities show a more uniform response across $(\theta, \phi)$ values than the imaging functionality, which means these functions are not highly dependent on incident angle. The imaging functionality is, by design, sensitive to the incident angle.

The red and green traces of Fig. \ref{Fig4}l show that as the wavelength shifts from 532 nm to 620 nm, the transmission through the 532 nm pixel smoothly decreases, and the transmission through the 620 nm pixel smoothly increases. These transmission values are averaged over both polarization and across all simulated incident angles $\phi \in [0, 360^\circ]$ and $\theta \in [0,5^\circ]$. As expected, the transmission to the 532 nm pixel is maximized for 532 nm wavelengths, and likewise for the 620 nm pixel. The traces smoothly vary between these two wavelengths, indicating that the power is predictably redistributed between the pixels in these cases. The solid blue line represents the average transmission to the \textit{correct} polarization pixel (for example, xz-polarized input being focused to the xz-polarization pixel), and the dashed blue line represents the average transmission to the \textit{incorrect} polarization pixel (for example, xz-polarized input being focused to the yz-polarization pixel). These transmission values are averaged over wavelength, and are averaged over incident angle in the same way the red and green traces are. While there is a drop in efficiency at the non-optimized wavelengths, there remains a high contrast between the solid and dashed blue lines at all wavelengths. 

The imaging functionality transmission values vary smoothly with respect to incident angles $(\theta,\phi)$, but this occurs in a less intuitive way than it would with a lens. Rather than the focus spot continuously shifting across the focal plane as $(\theta,\phi)$ varies, the focused spots instead brighten or dim while remaining approximately centered in the individual pixels. As an example, when $(\theta,\phi) = (2.5^{\circ},45^{\circ})$ the light is primarily scattered to the center of the $(0^{\circ},0^{\circ})$, $(5^{\circ},0^{\circ})$, $(5^{\circ},90^{\circ})$ pixels, as well as the relevant wavelength and polarization pixels. In this specific case 532 nm light is not focused to the 620 nm pixel which is the pixel that the incident planewave is oriented towards, and as such is where light would be focused if the device were replaced with a lens. Instead, the wavelength sorting and polarization sorting functionalities are mostly preserved under all excitation angles within an acceptance cone. Videos illustrating some examples as these parameters are swept are available in the Supporting Videos, along with a written description in the Supporting Information Section S6.

The effect of altering the polarization state is predictable since the polarization state of any input excitation can be described by two orthogonal linearly polarized states with a relative amplitude and phase shift between the orthogonal components. As an example, a simple experiment involving a linearly polarized source input into a Wollaston prism would show that as the source is rotated, the output of the Wollaston prism fluidly shifts from one linearly polarized output beam to the other. This behavior is observed in this device in the polarizing functionality, where the ratio between the polarizer pixels transmission can be used to infer the relative power of the two orthogonal linear polarization states. However, we would hope that the device performance for the wavelength and imaging functionalities is not adversely effected by the polarization states that were not explicitly optimized for. The efficiency of these functionalities may be altered as the cross-polarized output of one beam interferes with the parallel-polarized output of the orthogonally polarized input beam. We define the amount of cross-polarization by integrating the ratio $|E_x|^2/|E_y|^2$ under yz-polarized illumination over the output plane for all incident angles. The highest amount of cross-polarization is -11.9 dB at 620 nm and -13.6 dB at 532 nm. To verify that the various functionalities are not strongly affected, we analyze the results of all functionalities at all possible input polarization states by sweeping the relative amplitude and phase of the orthogonal input components, and we observe that interference effects can alter the transmission to the various pixels by a few percent. Data and further commentary on this matter is available in Supporting Information Section S1. Note that if a particular design seeks to further minimize cross-polarization, then cross-polarization can be explicitly minimized during the optimization. Doing so will consume some design degrees of freedom, which may detract from the efficiency of the various functionalities, but the optimization will not require more time since the number of electromagnetic simulations per iteration will be unchanged.

Based on these results the device could be used to analyze the state of incident light at wavelengths between 532 nm and 620 nm and inputs with incident angles within an approximately 5° cone. Additionally, the relative amplitude of the xz-polarized and yz-polarized component can be classified. The classification is done by measuring the ratio of intensity between pixels and using a look-up table to then approximate the state of light. In the case of classifying wavelength and imaging angle, the behavior of smoothly transitioning between states is non-trivial and is not exhibited in typical optical devices such as gratings and lenses. These components tend to spatially shift a beam in the focal plane as the wavelength is varied (grating) or the incident angle is varied (lens), whereas our device only alters the ratio of transmission to the various pixels with very little spatial shift of the beams. In the case of classifying polarization this behavior tends to occur naturally in other optical devices such as birefringent prisms because any polarization state can be decomposed into a coherent sum of two orthogonally polarized states. 

While classifying the wavelength and input angle states without ambiguity is not significantly affected by the incident polarization state, as was discussed in the Results section, it does require that the incident light be a monochromatic planewave. The planewave assumption is a common assumption employed in Shack-Hartman sensors and plenoptic sensors, and the assumption of monochromaticity can be satisfied with color filters, or if the light fundamentally comes from a narrow-band source such as a laser. Thus, there are numerous applications in which this device can be useful. The device can be tiled across a sensor array to enhance the functionality of an image sensor \cite{Camayd-Munoz2020,Hong2021,Zhao2021}. Such an array could be used to accurately classify the properties of a laser beam, including all fundamental properties of wavelength, polarization, and incident angle within the device’s acceptance cone. The angle-dependent nature of the device is similar in principle to angle-sensitive CMOS pixels\cite{Wang2009}, and could be used for lightfield imaging since the incident wavefront angle can be computationally determined using the relative intensity of the imaging pixels \cite{Adelson1992,Levoy2006,Levoy2009,Mait2018}. If coupled with a device such as a tunable bandpass filter, then a wavelength-dependent light field image can be obtained, with the added functionality of measuring the relative intensity of the two orthogonal linear polarization components for basic polarimetry. In general, we believe this type of device will open up new degrees of freedom in computational imaging applications.

\begin{figure}[htbp]
\centering\includegraphics[width=11cm]{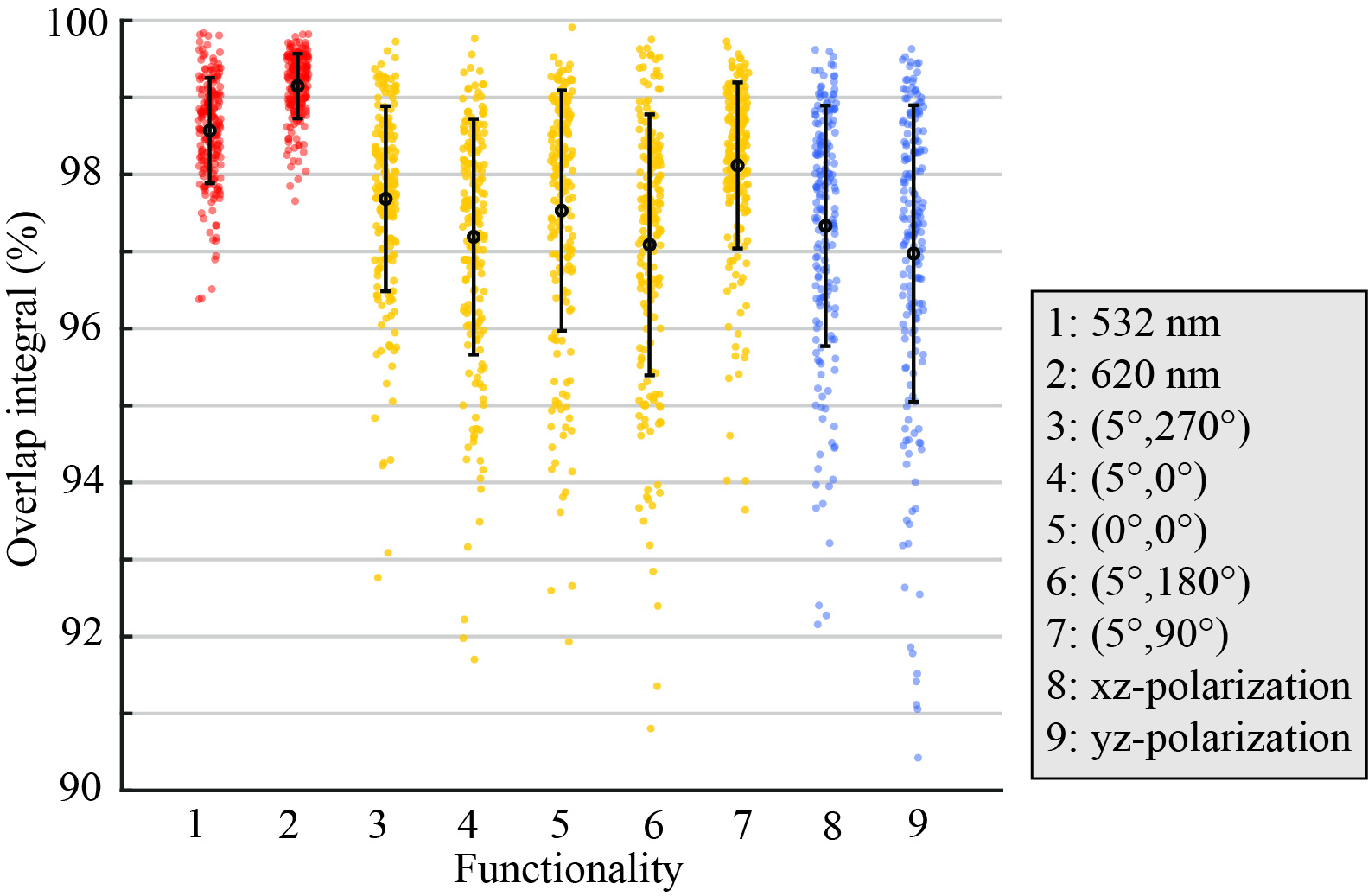}
\caption{The dependence of device performance on pixel distribution. For each functionality the overlap integral in Eq. \ref{eq:overlap} is evaluated on all combinations of pixel distributions. The average overlap integral is greater than 96.9\% for all functionalities, suggesting that the arbitrary choice of mapping functionalities to specific pixels has little effect on the behavior of the device.}
\label{fig5}
\end{figure}

One intriguing question we investigated is whether the qualitative behavior of the device, and its corresponding ability to interpolate input states effectively, depends on the specific assignment of functionalities to pixels. There are a large number of pixel permutations, and currently it is not easy to design a device for all of them. Instead, we investigated twenty different pixel distributions. Pixel distributions were chosen randomly, but were chosen to be sufficiently different from the original device by satisfying two criteria: 1) all pixels had to be moved from their original location, 2) no pixel could be rotated by $90^\circ$, $180^\circ$, or $270^\circ$ degrees, thus ensuring that devices were not similar by any rotational symmetry. Each device was optimized until the device was more than $15\%$ binary according to the following equation, where $B$ is the binarization,  $N$ is the total number of permittivity points in the design region, $\epsilon_i$ is the permittivity at the $i$th point, and $\epsilon_{min}$ and $\epsilon_{max}$ are the smallest and largest values of permittivity allowed in the design. 

\begin{gather}
    B = \frac{1}{N}\sum_i\left|\frac{\epsilon_i - \epsilon_{mid}}{\epsilon_{max}-\epsilon_{mid}}\right|, i\in \mathcal{D} \label{eq:binarization_definition}\\
    \epsilon_{mid} = \frac{\epsilon_{min}+\epsilon_{max}}{2}
\end{gather}

We compare the angular response of each functionality (the plots in Fig. \ref{Fig4}c-k) to one another using the overlap integral in Eq. \ref{eq:overlap}, integrated over all simulated $(\theta,\phi)$ points. This equation quantifies the similarity in the angular transmission profile, evaluating to zero if the responses are completely dissimilar and one if the responses are the same.

\begin{equation}
    \label{eq:overlap}
    \eta(T_1,T_2) = \left[ \frac{\left| \iint T_1 T_2 d\theta d\phi \right| ^2}{\iint |T_1|^2 d\theta d\phi \iint |T_2|^2 d\theta d\phi} \right] ^{1/2}
\end{equation}

For each functionality, we evaluate the overlap integral for all combinations of devices. Each result is plotted as a scatter point in Fig. \ref{fig5}, where the x-axis represents the nine different functionalities and the y-axis represents the computed overlap integral. The results indicate that all transmission profiles are very similar, suggesting that the qualitative behavior of the device is not strongly dependent on the mapping of the different functionalities to pixels. We withhold making any stronger claims than this qualitative similarity in behavior due to the limited number of pixel distributions that were studied.

\section{Discussion}

We presented a multi-functional 20-layer device made of SiO$_2$ and TiO$_2$ for use as a compressed sensing platform for the fundamental properties of an incident light wave: polarization, wavelength, and incident angle. The ability to classify wavelength and angle is contingent upon assumptions of the incident light. Monochromaticity is required for uniquely classifying the wavelength, which is a common assumption made in laser applications. The polarization functionality only measures the amplitudes of x- and y-polarized components, so a linear polarization is required for full classification. Future work could investigate introducing two additional polarization-classifying pixels to enable full-Stokes polarimetry. The last requirement is a locally linear phase front for classifying the incident angle, which is a commonly employed assumption used with Shack-Hartman and plenoptics sensors for applications such as wavefront sensing and computational imaging. This assumption is valid when the device is laterally small. Microlens arrays are often used for these applications, but the physical size of a microlens tends to be much larger than the devices presented here in order to span a sufficiently large number of sensor pixels.

The device was designed using topology optimization, aided by adjoint-based inverse design. The multi-objective optimization used techniques to approach a nearly binary solution during the continuous permittivity design phase, then concluded with a level-set approach that imposed fabrication constraints. It was found that the device functions at many more input conditions than those that were explicitly optimized for, which is a very promising feature that eases the computational burden of optimizing many different merit functions.

The fabrication of this device would be ambitious, but possibly achievable with state-of-the-art current  technology. The optimization procedure ensures that the device obeys a minimum feature size fabrication constraint of 50 nm, which is obtainable with optimized electron-beam lithography. A twenty-layer device with precise layer-to-layer alignment could be fabricated using similar processes used in CMOS foundries, although such processes are primarily used for electronics rather than optics and the materials used are not suitable for low-loss optical devices. A promising recent advancement involves designing a photonic device in a CMOS platform, then removing the metal layers with a wet etch to leave only materials made of the dielectric layers and air \cite{Fatemi2021}. This subtractive photonics platform could potentially be used to fabricate the devices presented here, particularly if the air was backfilled with TiO$_2$. In the near future, scaled-up versions of this device could be fabricated at infrared or terahertz frequencies to enable new imaging technologies.

\section*{Funding Sources}
This work was supported by the Defense Advanced Research Projects Agency
EXTREME program (HR00111720035), the Jet Propulsion Laboratory, California Institute of Technology,
Pasadena, CA, USA, under a contract with the National Aeronautics and Space Administration (PDRDF 107614-19AW0079), and Caltech Rothenberg Innovation Initiative. 

\bibliography{references}

\providecommand{\latin}[1]{#1}
\makeatletter
\providecommand{\doi}
  {\begingroup\let\do\@makeother\dospecials
  \catcode`\{=1 \catcode`\}=2 \doi@aux}
\providecommand{\doi@aux}[1]{\endgroup\texttt{#1}}
\makeatother
\providecommand*\mcitethebibliography{\thebibliography}
\csname @ifundefined\endcsname{endmcitethebibliography}
  {\let\endmcitethebibliography\endthebibliography}{}
\begin{mcitethebibliography}{36}
\providecommand*\natexlab[1]{#1}
\providecommand*\mciteSetBstSublistMode[1]{}
\providecommand*\mciteSetBstMaxWidthForm[2]{}
\providecommand*\mciteBstWouldAddEndPuncttrue
  {\def\EndOfBibitem{\unskip.}}
\providecommand*\mciteBstWouldAddEndPunctfalse
  {\let\EndOfBibitem\relax}
\providecommand*\mciteSetBstMidEndSepPunct[3]{}
\providecommand*\mciteSetBstSublistLabelBeginEnd[3]{}
\providecommand*\EndOfBibitem{}
\mciteSetBstSublistMode{f}
\mciteSetBstMaxWidthForm{subitem}{(\alph{mcitesubitemcount})}
\mciteSetBstSublistLabelBeginEnd
  {\mcitemaxwidthsubitemform\space}
  {\relax}
  {\relax}

\bibitem[Süzen \latin{et~al.}(2010)Süzen, Giannoula, and Durduran]{Suzen2010}
Süzen,~M.; Giannoula,~A.; Durduran,~T. Compressed sensing in diffuse optical
  tomography. \emph{Opt. Express} \textbf{2010}, \emph{18}, 23676--23690\relax
\mciteBstWouldAddEndPuncttrue
\mciteSetBstMidEndSepPunct{\mcitedefaultmidpunct}
{\mcitedefaultendpunct}{\mcitedefaultseppunct}\relax
\EndOfBibitem
\bibitem[Pégard \latin{et~al.}(2016)Pégard, Liu, Antipa, Gerlock, Adesnik,
  and Waller]{Pegard2016}
Pégard,~N.~C.; Liu,~H.-Y.; Antipa,~N.; Gerlock,~M.; Adesnik,~H.; Waller,~L.
  Compressive light-field microscopy for 3D neural activity recording.
  \emph{Optica} \textbf{2016}, \emph{3}, 517--524\relax
\mciteBstWouldAddEndPuncttrue
\mciteSetBstMidEndSepPunct{\mcitedefaultmidpunct}
{\mcitedefaultendpunct}{\mcitedefaultseppunct}\relax
\EndOfBibitem
\bibitem[Mait \latin{et~al.}(2018)Mait, Euliss, and Athale]{Mait2018}
Mait,~J.~N.; Euliss,~G.~W.; Athale,~R.~A. {Computational imaging}.
  \emph{Advances in Optics and Photonics} \textbf{2018}, \emph{10}, 409\relax
\mciteBstWouldAddEndPuncttrue
\mciteSetBstMidEndSepPunct{\mcitedefaultmidpunct}
{\mcitedefaultendpunct}{\mcitedefaultseppunct}\relax
\EndOfBibitem
\bibitem[Antipa \latin{et~al.}(2019)Antipa, Oare, Bostan, Ng, and
  Waller]{https://doi.org/10.48550/arxiv.1905.13221}
Antipa,~N.; Oare,~P.; Bostan,~E.; Ng,~R.; Waller,~L. Video from Stills:
  Lensless Imaging with Rolling Shutter. 2019;
  \url{https://arxiv.org/abs/1905.13221}\relax
\mciteBstWouldAddEndPuncttrue
\mciteSetBstMidEndSepPunct{\mcitedefaultmidpunct}
{\mcitedefaultendpunct}{\mcitedefaultseppunct}\relax
\EndOfBibitem
\bibitem[Roberts \latin{et~al.}(2022)Roberts, Ballew, Zheng, Garcia,
  Camayd-Muñoz, Hon, and Faraon]{Roberts2022}
Roberts,~G.; Ballew,~C.; Zheng,~T.; Garcia,~J.~C.; Camayd-Muñoz,~S.; Hon,~P.
  W.~C.; Faraon,~A. 3D-Patterned Inverse-Designed Mid-Infrared Metaoptics.
  2022; \url{https://arxiv.org/abs/2209.07553}\relax
\mciteBstWouldAddEndPuncttrue
\mciteSetBstMidEndSepPunct{\mcitedefaultmidpunct}
{\mcitedefaultendpunct}{\mcitedefaultseppunct}\relax
\EndOfBibitem
\bibitem[Yu and Capasso(2014)Yu, and Capasso]{Yu2014}
Yu,~N.; Capasso,~F. {Flat optics with designer metasurfaces}. 2014;
  \url{www.nature.com/naturematerials}\relax
\mciteBstWouldAddEndPuncttrue
\mciteSetBstMidEndSepPunct{\mcitedefaultmidpunct}
{\mcitedefaultendpunct}{\mcitedefaultseppunct}\relax
\EndOfBibitem
\bibitem[Arbabi \latin{et~al.}(2015)Arbabi, Horie, Bagheri, and
  Faraon]{Arbabi2015}
Arbabi,~A.; Horie,~Y.; Bagheri,~M.; Faraon,~A. {Dielectric metasurfaces for
  complete control of phase and polarization with subwavelength spatial
  resolution and high transmission}. \emph{Nature Nanotechnology}
  \textbf{2015}, \emph{10}, 937--943\relax
\mciteBstWouldAddEndPuncttrue
\mciteSetBstMidEndSepPunct{\mcitedefaultmidpunct}
{\mcitedefaultendpunct}{\mcitedefaultseppunct}\relax
\EndOfBibitem
\bibitem[Chen \latin{et~al.}(2016)Chen, Taylor, and Yu]{Chen2016}
Chen,~H.~T.; Taylor,~A.~J.; Yu,~N. A review of metasurfaces: Physics and
  applications. \emph{Reports on Progress in Physics} \textbf{2016},
  \emph{79}\relax
\mciteBstWouldAddEndPuncttrue
\mciteSetBstMidEndSepPunct{\mcitedefaultmidpunct}
{\mcitedefaultendpunct}{\mcitedefaultseppunct}\relax
\EndOfBibitem
\bibitem[Miller(2007)]{Miller2007}
Miller,~D. A.~B. Fundamental limit for optical components. \emph{Journal of the
  Optical Society of America B} \textbf{2007}, \emph{24}, A1\relax
\mciteBstWouldAddEndPuncttrue
\mciteSetBstMidEndSepPunct{\mcitedefaultmidpunct}
{\mcitedefaultendpunct}{\mcitedefaultseppunct}\relax
\EndOfBibitem
\bibitem[Arbabi and Faraon(2017)Arbabi, and Faraon]{Arbabi2017}
Arbabi,~A.; Faraon,~A. {Fundamental limits of ultrathin metasurfaces}.
  \emph{Scientific Reports} \textbf{2017}, \emph{7}, 1--9\relax
\mciteBstWouldAddEndPuncttrue
\mciteSetBstMidEndSepPunct{\mcitedefaultmidpunct}
{\mcitedefaultendpunct}{\mcitedefaultseppunct}\relax
\EndOfBibitem
\bibitem[Pfeiffer and Grbic(2013)Pfeiffer, and Grbic]{Pfeiffer2013}
Pfeiffer,~C.; Grbic,~A. Cascaded metasurfaces for complete phase and
  polarization control. \emph{Applied Physics Letters} \textbf{2013},
  \emph{102}, 231116\relax
\mciteBstWouldAddEndPuncttrue
\mciteSetBstMidEndSepPunct{\mcitedefaultmidpunct}
{\mcitedefaultendpunct}{\mcitedefaultseppunct}\relax
\EndOfBibitem
\bibitem[Avayu \latin{et~al.}(2017)Avayu, Almeida, Prior, and
  Ellenbogen]{Avayu2017}
Avayu,~O.; Almeida,~E.; Prior,~Y.; Ellenbogen,~T. Composite functional
  metasurfaces for multispectral achromatic optics. \emph{Nature
  Communications} \textbf{2017}, \emph{8}, 14992\relax
\mciteBstWouldAddEndPuncttrue
\mciteSetBstMidEndSepPunct{\mcitedefaultmidpunct}
{\mcitedefaultendpunct}{\mcitedefaultseppunct}\relax
\EndOfBibitem
\bibitem[Zhou \latin{et~al.}(2018)Zhou, Kravchenko, Wang, Nolen, Gu, and
  Valentine]{Zhou2018}
Zhou,~Y.; Kravchenko,~I.~I.; Wang,~H.; Nolen,~J.~R.; Gu,~G.; Valentine,~J.
  Multilayer Noninteracting Dielectric Metasurfaces for Multiwavelength
  Metaoptics. \emph{Nano Letters} \textbf{2018}, \emph{18}, 7529--7537, PMID:
  30394751\relax
\mciteBstWouldAddEndPuncttrue
\mciteSetBstMidEndSepPunct{\mcitedefaultmidpunct}
{\mcitedefaultendpunct}{\mcitedefaultseppunct}\relax
\EndOfBibitem
\bibitem[Huang \latin{et~al.}(2019)Huang, Rubin, Ambrosio, Shi, Devlin, Qiu,
  and Capasso]{Huang2019}
Huang,~Y.-W.; Rubin,~N.~A.; Ambrosio,~A.; Shi,~Z.; Devlin,~R.~C.; Qiu,~C.-W.;
  Capasso,~F. Versatile total angular momentum generation using cascaded
  J-plates. \emph{Opt. Express} \textbf{2019}, \emph{27}, 7469--7484\relax
\mciteBstWouldAddEndPuncttrue
\mciteSetBstMidEndSepPunct{\mcitedefaultmidpunct}
{\mcitedefaultendpunct}{\mcitedefaultseppunct}\relax
\EndOfBibitem
\bibitem[Mansouree \latin{et~al.}(2020)Mansouree, Kwon, Arbabi, McClung,
  Faraon, and Arbabi]{Mansouree2020}
Mansouree,~M.; Kwon,~H.; Arbabi,~E.; McClung,~A.; Faraon,~A.; Arbabi,~A.
  {Multifunctional 2.5D metastructures enabled by adjoint optimization}.
  \emph{Optica} \textbf{2020}, \emph{7}, 77\relax
\mciteBstWouldAddEndPuncttrue
\mciteSetBstMidEndSepPunct{\mcitedefaultmidpunct}
{\mcitedefaultendpunct}{\mcitedefaultseppunct}\relax
\EndOfBibitem
\bibitem[Kamali \latin{et~al.}(2018)Kamali, Arbabi, Arbabi, and
  Faraon]{Kamali2018}
Kamali,~S.~M.; Arbabi,~E.; Arbabi,~A.; Faraon,~A. {A review of dielectric
  optical metasurfaces for wavefront control}. 2018;
  \url{https://doi.org/10.1515/nanoph-2017-0129}\relax
\mciteBstWouldAddEndPuncttrue
\mciteSetBstMidEndSepPunct{\mcitedefaultmidpunct}
{\mcitedefaultendpunct}{\mcitedefaultseppunct}\relax
\EndOfBibitem
\bibitem[Miller(2023)]{Miller2023}
Miller,~D. A.~B. Why optics needs thickness. \emph{Science} \textbf{2023},
  \emph{379}, 41--45\relax
\mciteBstWouldAddEndPuncttrue
\mciteSetBstMidEndSepPunct{\mcitedefaultmidpunct}
{\mcitedefaultendpunct}{\mcitedefaultseppunct}\relax
\EndOfBibitem
\bibitem[Miller(2012)]{Miller2012}
Miller,~O. Photonic Design: From Fundamental Solar Cell Physics to
  Computational Inverse Design. Ph.D.\ thesis, EECS Department, University of
  California, Berkeley, 2012\relax
\mciteBstWouldAddEndPuncttrue
\mciteSetBstMidEndSepPunct{\mcitedefaultmidpunct}
{\mcitedefaultendpunct}{\mcitedefaultseppunct}\relax
\EndOfBibitem
\bibitem[Lalau-Keraly \latin{et~al.}(2013)Lalau-Keraly, Bhargava, Miller, and
  Yablonovitch]{Lalau-Keraly2013}
Lalau-Keraly,~C.~M.; Bhargava,~S.; Miller,~O.~D.; Yablonovitch,~E. {Adjoint
  shape optimization applied to electromagnetic design}. \emph{Optics Express}
  \textbf{2013}, \emph{21}, 21693\relax
\mciteBstWouldAddEndPuncttrue
\mciteSetBstMidEndSepPunct{\mcitedefaultmidpunct}
{\mcitedefaultendpunct}{\mcitedefaultseppunct}\relax
\EndOfBibitem
\bibitem[Molesky \latin{et~al.}(2018)Molesky, Lin, Piggott, Jin, Vucković, and
  Rodriguez]{Molesky2018}
Molesky,~S.; Lin,~Z.; Piggott,~A.~Y.; Jin,~W.; Vucković,~J.; Rodriguez,~A.~W.
  Inverse design in nanophotonics. \emph{Nature Photonics} \textbf{2018},
  \emph{12}, 659--670\relax
\mciteBstWouldAddEndPuncttrue
\mciteSetBstMidEndSepPunct{\mcitedefaultmidpunct}
{\mcitedefaultendpunct}{\mcitedefaultseppunct}\relax
\EndOfBibitem
\bibitem[Jensen and Sigmund(2004)Jensen, and Sigmund]{Jensen2004}
Jensen,~J.~S.; Sigmund,~O. {Systematic design of photonic crystal structures
  using topology optimization: Low-loss waveguide bends}. \emph{Applied Physics
  Letters} \textbf{2004}, \emph{84}, 2022--2024\relax
\mciteBstWouldAddEndPuncttrue
\mciteSetBstMidEndSepPunct{\mcitedefaultmidpunct}
{\mcitedefaultendpunct}{\mcitedefaultseppunct}\relax
\EndOfBibitem
\bibitem[Jensen and Sigmund(2011)Jensen, and Sigmund]{Jensen2011}
Jensen,~J.~S.; Sigmund,~O. {Topology optimization for nano-photonics}. 2011;
  \url{https://onlinelibrary.wiley.com/doi/full/10.1002/lpor.201000014
  https://onlinelibrary.wiley.com/doi/abs/10.1002/lpor.201000014
  https://onlinelibrary.wiley.com/doi/10.1002/lpor.201000014}\relax
\mciteBstWouldAddEndPuncttrue
\mciteSetBstMidEndSepPunct{\mcitedefaultmidpunct}
{\mcitedefaultendpunct}{\mcitedefaultseppunct}\relax
\EndOfBibitem
\bibitem[Piggott \latin{et~al.}(2015)Piggott, Lu, Lagoudakis, Petykiewicz,
  Babinec, and Vuckovi{\'{c}}]{Piggott2015}
Piggott,~A.~Y.; Lu,~J.; Lagoudakis,~K.~G.; Petykiewicz,~J.; Babinec,~T.~M.;
  Vuckovi{\'{c}},~J. {Inverse design and demonstration of a compact and
  broadband on-chip wavelength demultiplexer}. \emph{Nature Photonics}
  \textbf{2015}, \emph{9}, 374--377\relax
\mciteBstWouldAddEndPuncttrue
\mciteSetBstMidEndSepPunct{\mcitedefaultmidpunct}
{\mcitedefaultendpunct}{\mcitedefaultseppunct}\relax
\EndOfBibitem
\bibitem[Camayd-Mu{\~{n}}oz \latin{et~al.}(2020)Camayd-Mu{\~{n}}oz, Ballew,
  Roberts, and Faraon]{Camayd-Munoz2020}
Camayd-Mu{\~{n}}oz,~P.; Ballew,~C.; Roberts,~G.; Faraon,~A. {Multifunctional
  volumetric meta-optics for color and polarization image sensors}.
  \emph{Optica} \textbf{2020}, \emph{7}, 280\relax
\mciteBstWouldAddEndPuncttrue
\mciteSetBstMidEndSepPunct{\mcitedefaultmidpunct}
{\mcitedefaultendpunct}{\mcitedefaultseppunct}\relax
\EndOfBibitem
\bibitem[Ballew \latin{et~al.}(2021)Ballew, Roberts, Camayd-Mu{\~{n}}oz,
  Debbas, and Faraon]{Ballew2021}
Ballew,~C.; Roberts,~G.; Camayd-Mu{\~{n}}oz,~S.; Debbas,~M.~F.; Faraon,~A.
  {Mechanically reconfigurable multi-functional meta-optics studied at
  microwave frequencies}. \emph{Scientific Reports} \textbf{2021}, \emph{11},
  11145\relax
\mciteBstWouldAddEndPuncttrue
\mciteSetBstMidEndSepPunct{\mcitedefaultmidpunct}
{\mcitedefaultendpunct}{\mcitedefaultseppunct}\relax
\EndOfBibitem
\bibitem[Vercruysse \latin{et~al.}(2019)Vercruysse, Sapra, Su, Trivedi, and
  Vu{\v{c}}kovi{\'{c}}]{Vercruysse2019}
Vercruysse,~D.; Sapra,~N.~V.; Su,~L.; Trivedi,~R.; Vu{\v{c}}kovi{\'{c}},~J.
  {Analytical level set fabrication constraints for inverse design}.
  \emph{Scientific Reports} \textbf{2019}, \emph{9}, 1--7\relax
\mciteBstWouldAddEndPuncttrue
\mciteSetBstMidEndSepPunct{\mcitedefaultmidpunct}
{\mcitedefaultendpunct}{\mcitedefaultseppunct}\relax
\EndOfBibitem
\bibitem[Ballew \latin{et~al.}(0)Ballew, Roberts, Zheng, and
  Faraon]{Ballew2022}
Ballew,~C.; Roberts,~G.; Zheng,~T.; Faraon,~A. Constraining Continuous Topology
  Optimizations to Discrete Solutions for Photonic Applications. 0;
  \url{https://doi.org/10.1021/acsphotonics.2c00862}\relax
\mciteBstWouldAddEndPuncttrue
\mciteSetBstMidEndSepPunct{\mcitedefaultmidpunct}
{\mcitedefaultendpunct}{\mcitedefaultseppunct}\relax
\EndOfBibitem
\bibitem[Lebbe \latin{et~al.}(2019)Lebbe, Dapogny, Oudet, Hassan, and
  Gliere]{Lebbe2019}
Lebbe,~N.; Dapogny,~C.; Oudet,~E.; Hassan,~K.; Gliere,~A. {Robust shape and
  topology optimization of nanophotonic devices using the level set method}.
  \emph{Journal of Computational Physics} \textbf{2019}, \emph{395},
  710--746\relax
\mciteBstWouldAddEndPuncttrue
\mciteSetBstMidEndSepPunct{\mcitedefaultmidpunct}
{\mcitedefaultendpunct}{\mcitedefaultseppunct}\relax
\EndOfBibitem
\bibitem[Hong \latin{et~al.}(2021)Hong, Son, Kim, Mun, Sung, and Lee]{Hong2021}
Hong,~J.; Son,~H.; Kim,~C.; Mun,~S.-E.; Sung,~J.; Lee,~B. {Absorptive
  metasurface color filters based on hyperbolic metamaterials for a CMOS image
  sensor}. \emph{Opt. Express} \textbf{2021}, \emph{29}, 3643--3658\relax
\mciteBstWouldAddEndPuncttrue
\mciteSetBstMidEndSepPunct{\mcitedefaultmidpunct}
{\mcitedefaultendpunct}{\mcitedefaultseppunct}\relax
\EndOfBibitem
\bibitem[Zhao \latin{et~al.}(2021)Zhao, Catrysse, and Fan]{Zhao2021}
Zhao,~N.; Catrysse,~P.~B.; Fan,~S. {Perfect RGB-IR Color Routers for
  Sub-Wavelength Size CMOS Image Sensor Pixels}. \emph{Advanced Photonics
  Research} \textbf{2021}, \emph{2}, 2000048\relax
\mciteBstWouldAddEndPuncttrue
\mciteSetBstMidEndSepPunct{\mcitedefaultmidpunct}
{\mcitedefaultendpunct}{\mcitedefaultseppunct}\relax
\EndOfBibitem
\bibitem[Wang \latin{et~al.}(2009)Wang, Gill, and Molnar]{Wang2009}
Wang,~A.; Gill,~P.; Molnar,~A. {Light field image sensors based on the Talbot
  effect}. \emph{Appl. Opt.} \textbf{2009}, \emph{48}, 5897--5905\relax
\mciteBstWouldAddEndPuncttrue
\mciteSetBstMidEndSepPunct{\mcitedefaultmidpunct}
{\mcitedefaultendpunct}{\mcitedefaultseppunct}\relax
\EndOfBibitem
\bibitem[Adelson and Wang(1992)Adelson, and Wang]{Adelson1992}
Adelson,~E.~H.; Wang,~J. Y.~A. {Single lens stereo with a plenoptic camera}.
  \emph{IEEE Transactions on Pattern Analysis and Machine Intelligence}
  \textbf{1992}, \emph{14}, 99--106\relax
\mciteBstWouldAddEndPuncttrue
\mciteSetBstMidEndSepPunct{\mcitedefaultmidpunct}
{\mcitedefaultendpunct}{\mcitedefaultseppunct}\relax
\EndOfBibitem
\bibitem[Levoy \latin{et~al.}(2006)Levoy, Ng, Adams, Footer, and
  Horowitz]{Levoy2006}
Levoy,~M.; Ng,~R.; Adams,~A.; Footer,~M.; Horowitz,~M. {Light Field
  Microscopy}. \emph{ACM Trans. Graph.} \textbf{2006}, \emph{25},
  924--934\relax
\mciteBstWouldAddEndPuncttrue
\mciteSetBstMidEndSepPunct{\mcitedefaultmidpunct}
{\mcitedefaultendpunct}{\mcitedefaultseppunct}\relax
\EndOfBibitem
\bibitem[Levoy \latin{et~al.}(2009)Levoy, Zhang, and McDowall]{Levoy2009}
Levoy,~M.; Zhang,~Z.; McDowall,~I. {Recording and controlling the 4D light
  field in a microscope using microlens arrays}. \emph{Journal of Microscopy}
  \textbf{2009}, \emph{235}, 144--162\relax
\mciteBstWouldAddEndPuncttrue
\mciteSetBstMidEndSepPunct{\mcitedefaultmidpunct}
{\mcitedefaultendpunct}{\mcitedefaultseppunct}\relax
\EndOfBibitem
\bibitem[Fatemi \latin{et~al.}(2021)Fatemi, Ives, Khachaturian, and
  Hajimiri]{Fatemi2021}
Fatemi,~R.; Ives,~C.; Khachaturian,~A.; Hajimiri,~A. {Subtractive photonics}.
  \emph{Opt. Express} \textbf{2021}, \emph{29}, 877--893\relax
\mciteBstWouldAddEndPuncttrue
\mciteSetBstMidEndSepPunct{\mcitedefaultmidpunct}
{\mcitedefaultendpunct}{\mcitedefaultseppunct}\relax
\EndOfBibitem
\end{mcitethebibliography}


\providecommand{\latin}[1]{#1}
\makeatletter
\providecommand{\doi}
  {\begingroup\let\do\@makeother\dospecials
  \catcode`\{=1 \catcode`\}=2 \doi@aux}
\providecommand{\doi@aux}[1]{\endgroup\texttt{#1}}
\makeatother
\providecommand*\mcitethebibliography{\thebibliography}
\csname @ifundefined\endcsname{endmcitethebibliography}
  {\let\endmcitethebibliography\endthebibliography}{}
\begin{mcitethebibliography}{3}
\providecommand*\natexlab[1]{#1}
\providecommand*\mciteSetBstSublistMode[1]{}
\providecommand*\mciteSetBstMaxWidthForm[2]{}
\providecommand*\mciteBstWouldAddEndPuncttrue
  {\def\EndOfBibitem{\unskip.}}
\providecommand*\mciteBstWouldAddEndPunctfalse
  {\let\EndOfBibitem\relax}
\providecommand*\mciteSetBstMidEndSepPunct[3]{}
\providecommand*\mciteSetBstSublistLabelBeginEnd[3]{}
\providecommand*\EndOfBibitem{}
\mciteSetBstSublistMode{f}
\mciteSetBstMaxWidthForm{subitem}{(\alph{mcitesubitemcount})}
\mciteSetBstSublistLabelBeginEnd
  {\mcitemaxwidthsubitemform\space}
  {\relax}
  {\relax}

\bibitem[Vercruysse \latin{et~al.}(2019)Vercruysse, Sapra, Su, Trivedi, and
  Vu{\v{c}}kovi{\'{c}}]{Vercruysse2019}
Vercruysse,~D.; Sapra,~N.~V.; Su,~L.; Trivedi,~R.; Vu{\v{c}}kovi{\'{c}},~J.
  {Analytical level set fabrication constraints for inverse design}.
  \emph{Scientific Reports} \textbf{2019}, \emph{9}, 1--7\relax
\mciteBstWouldAddEndPuncttrue
\mciteSetBstMidEndSepPunct{\mcitedefaultmidpunct}
{\mcitedefaultendpunct}{\mcitedefaultseppunct}\relax
\EndOfBibitem
\bibitem[Ballew \latin{et~al.}(2021)Ballew, Roberts, Camayd-Mu{\~{n}}oz,
  Debbas, and Faraon]{Ballew2021}
Ballew,~C.; Roberts,~G.; Camayd-Mu{\~{n}}oz,~S.; Debbas,~M.~F.; Faraon,~A.
  {Mechanically reconfigurable multi-functional meta-optics studied at
  microwave frequencies}. \emph{Scientific Reports} \textbf{2021}, \emph{11},
  11145\relax
\mciteBstWouldAddEndPuncttrue
\mciteSetBstMidEndSepPunct{\mcitedefaultmidpunct}
{\mcitedefaultendpunct}{\mcitedefaultseppunct}\relax
\EndOfBibitem
\end{mcitethebibliography}

\end{document}

\maketitle

\section{Effects of cross-polarization}

The device discussed in the main manuscript sorts incident light based on its polarization, wavelength, and incident angle. The device was designed explicitly for two orthogonal linearly-polarized sources, one polarized in the $xz$ plane and the other in the $yz$ plane. Here, we wish to determine the effect on device performance when excited with planewaves of arbitrary polarization. The response of the device for any polarization state can be determined by coherently summing the electric fields, applying relative amplitude and phase shifts (i.e. a Jones vector) to the two linearly-polarized components to obtain arbitrary elliptical polarizations.

The Jones vector is defined as:

\begin{equation*}
\begin{pmatrix}
E_{0x}\\
E_{0y}e^{i\phi_J}
\end{pmatrix}
\end{equation*}

To sweep the Jones vector over all possibilities, we introduce a variable $\theta_J \in [0,\pi]$ and set $E_{0x}=\cos(\theta_J)$ and $E_{0y}=\sin(\theta_J)$. We then plot the transmission to all pixels as function of $(\theta_J, \phi_J)$ with $\phi_J \in [0,\pi]$. Since we wish to observe interference effects here, we do not average over any properties as was done in quantifying performance in the main manuscript. Instead, the results for all training modes are plotted independently. The transmission to the pixels do exhibit a dependence on the incident angle, but it is not substantial enough to alter the behavior of the device. Thus, the device will exhibit the behavior reported in the main manuscript under any polarization input.

Figures \ref{FigS1}-\ref{FigS5} below show these results. $\theta_J$, the relative amplitude, is the x-axis and $\phi_J$, the relative phase, is the y-axis for each subplot. Each figure contains the results for all pixels at a particular incident angle $(\theta,\phi)$. The specific pixels to which transmission is \textit{not} maximized are faded. The ordering of the pixels is identical to that of Fig. 1 in the main text.

\begin{figure}[htbp]
\centerline{\includegraphics[width=15cm]{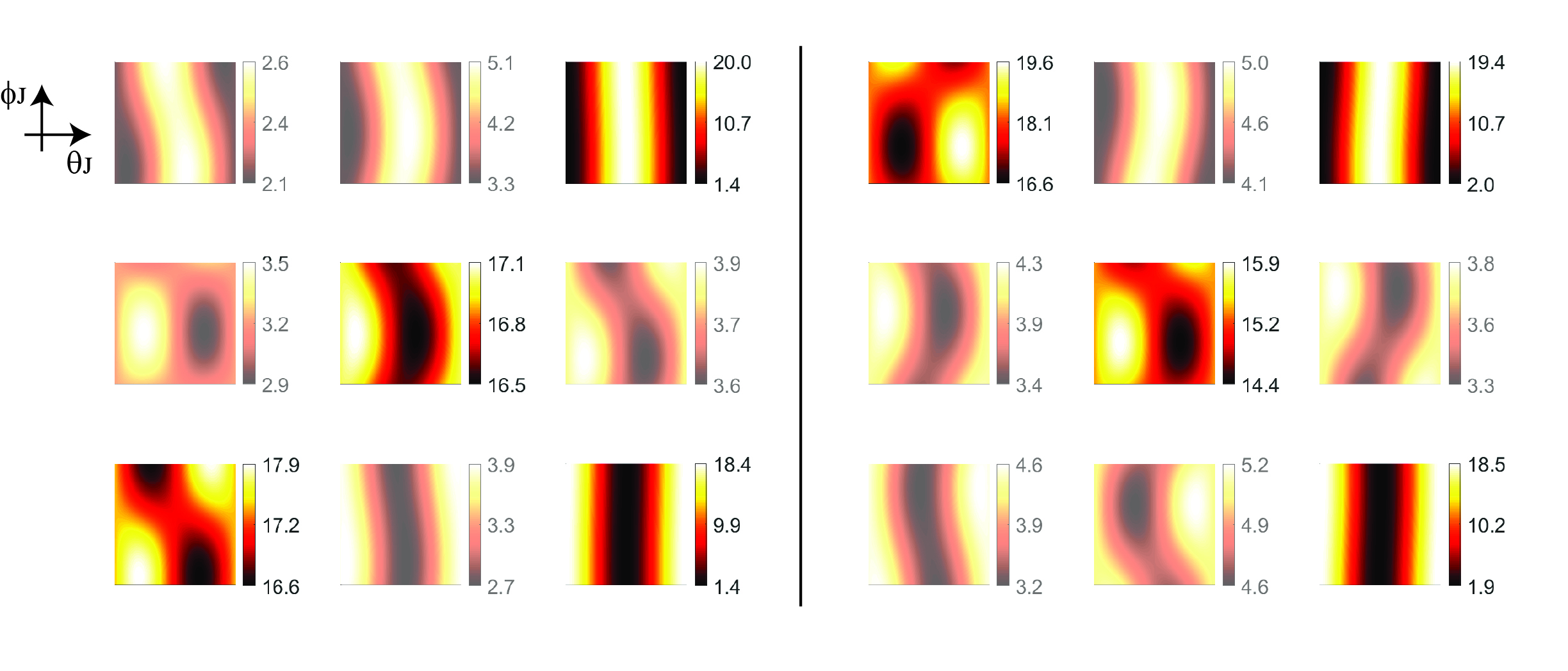}}
\vspace*{-8mm}\caption{ $(\theta,\phi) = (0^\circ, 0^\circ)$. Left set of figures correspond to 620 nm, the right set to 532 nm. } 
\label{FigS1}
\end{figure}

\begin{figure}[htbp]
\centerline{\includegraphics[width=15cm]{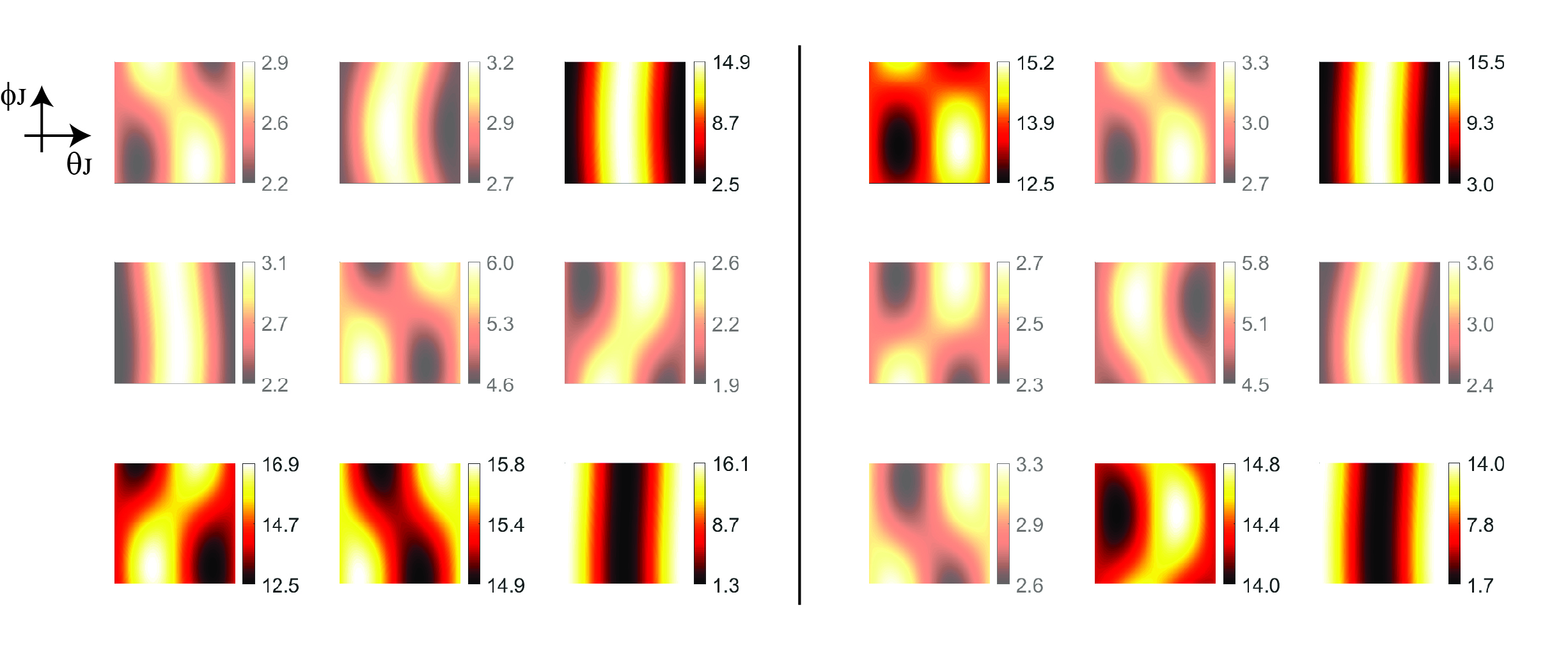}}
\vspace*{-8mm}\caption{ $(\theta,\phi) = (5^\circ, 0^\circ)$. Left set of figures correspond to 620 nm, the right set to 532 nm. } 
\label{FigS2}
\end{figure}

\begin{figure}[htbp]
\centerline{\includegraphics[width=15cm]{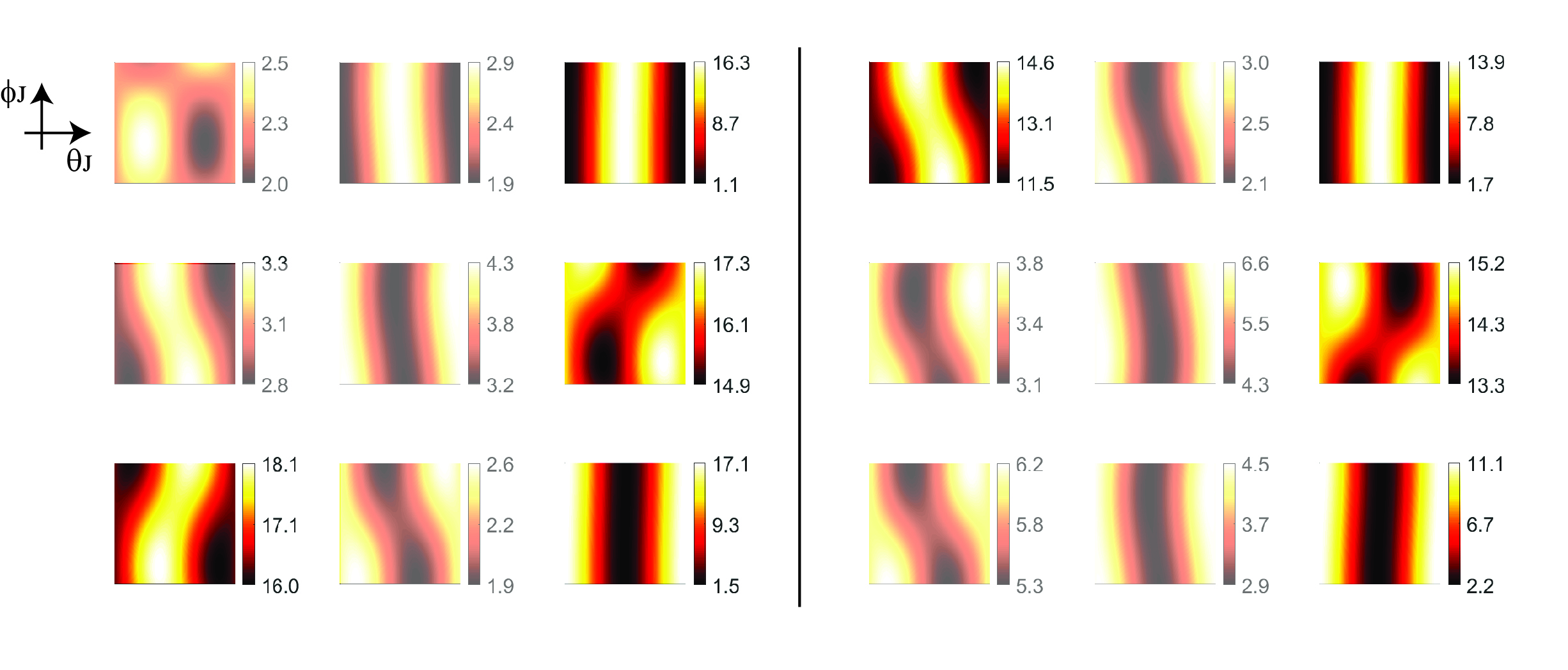}}
\vspace*{-8mm}\caption{ $(\theta,\phi) = (5^\circ, 90^\circ)$. Left set of figures correspond to 620 nm, the right set to 532 nm. } 
\label{FigS3}
\end{figure}

\begin{figure}[htbp]
\centerline{\includegraphics[width=15cm]{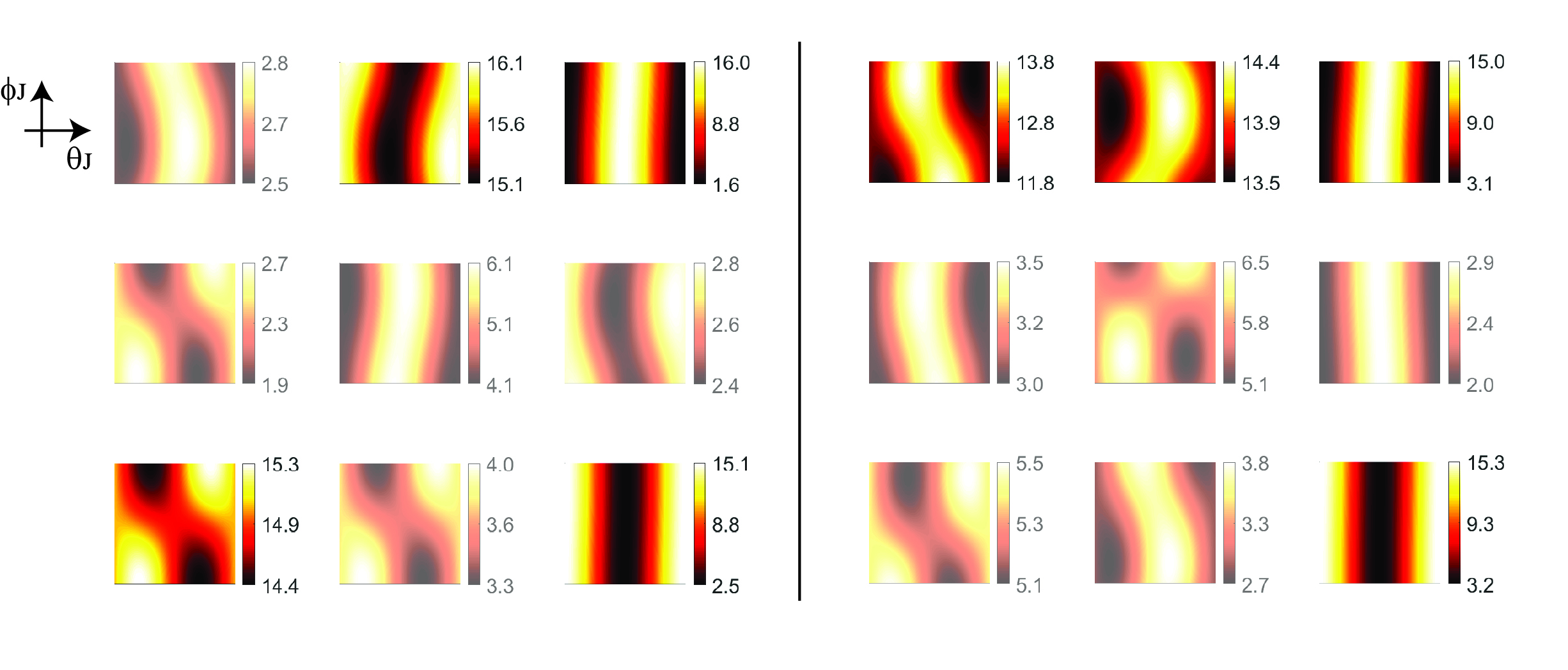}}
\vspace*{-8mm}\caption{ $(\theta,\phi) = (5^\circ, 180^\circ)$. Left set of figures correspond to 620 nm, the right set to 532 nm. } 
\label{FigS4}
\end{figure}

\begin{figure}[htbp]
\centerline{\includegraphics[width=15cm]{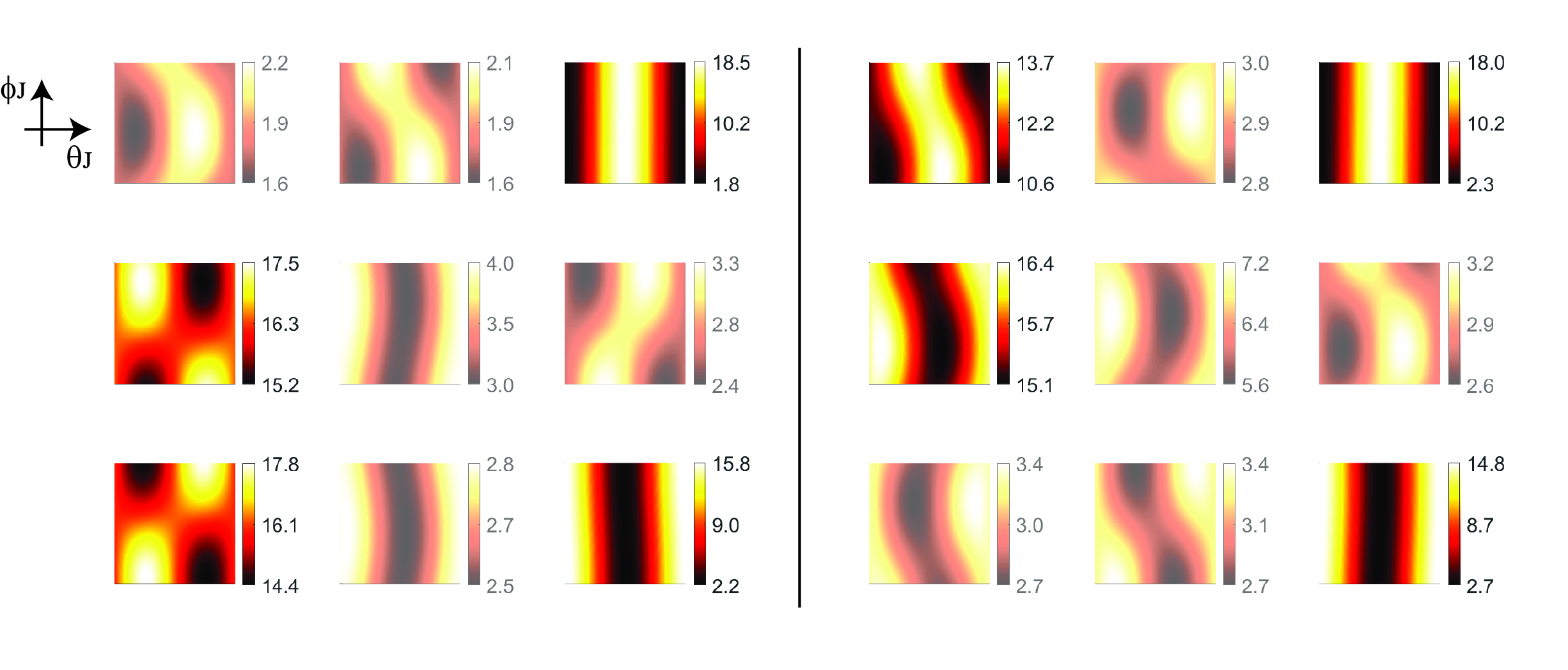}}
\vspace*{-8mm}\caption{ $(\theta,\phi) = (5^\circ, 270^\circ)$. Left set of figures correspond to 620 nm, the right set to 532 nm. } 
\label{FigS5}
\end{figure}

\pagebreak

\section{Level-set optimization}
The term level-set optimization stems from treating the device boundaries as the zero-level contour of a level-set function $\phi(x,y)$. The level-set function (LSF) is perturbed in accordance with the gradient, which has the effect of perturbing the zero-level contour of the LSF and the associated device boundaries to increase performance. We use a signed-distance function to define the level-set function (LSF), where the value of $\phi(x,y)$ is proportional to the signed distance from the device boundary. The gradients of the electromagnetic FoM and a fabrication penalty function are combined and used to perturb the LSF, which has the effect of perturbing the boundary of the device such that the electromagnetic FoM is increased and the fabrication penalty term is decreased. The fabrication penalty term includes both a minimum radius of curvature constraint and a minimum gap size constraint of 60 nm. The LSF is then recomputed to ensure it remains a signed-distance function. This process is repeated until the FoM has converged (recovering the performance lost from the binarization step), and the fabrication penalty term is minimized.

It is critical that we preserve fabrication restrictions during the level-set optimization, and we do so here using the techniques described in detail in Ref. \cite{Vercruysse2019}. To briefly summarize that work, a multi-dimensional fabrication penalty function is first analytically computed over the entire device region. This function includes two types of fabrication constraints: one limits the radius of curvature of the device boundaries, and the other limits the smallest gap size of the device. The terms are integrated over the entire design region, yielding a real number (the fabrication penalty term) that we wish to minimize. The gradient of this fabrication penalty term is computed over the device region using a finite-difference approximation and is subsequently combined with the gradient of the electromagnetic FoM that is computed through the adjoint method. The level-set function is perturbed in the direction of this combined gradient, resulting in a shifting of the material boundaries that co-optimizes the electromagnetic FoM and the fabrication penalty term.

\section{Convergence plots}

\begin{figure}[htbp]
\centering\includegraphics[width=10cm]{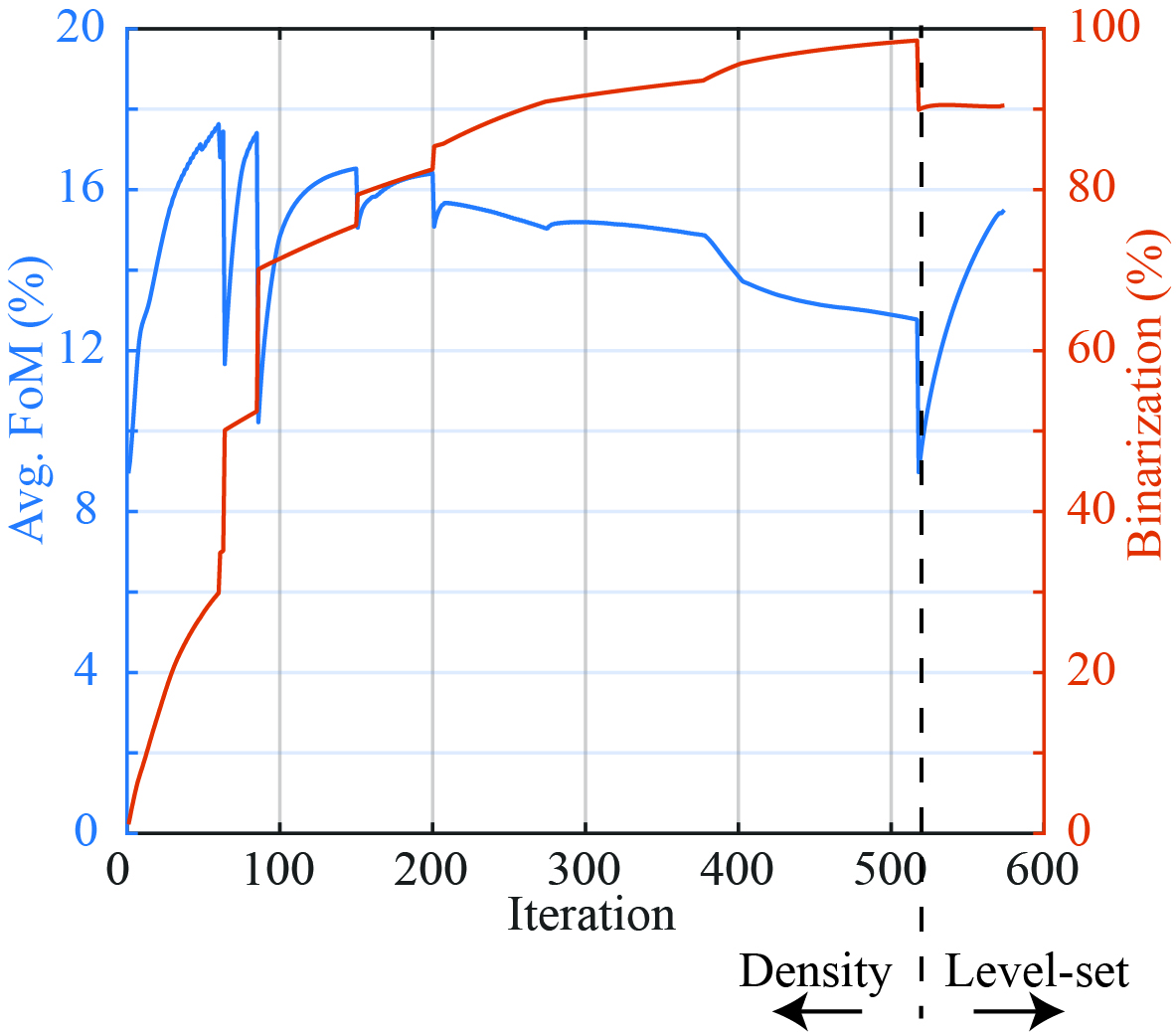}
\caption{\textbf{Convergence plots of the 3D metaoptics device.} The initial density-based optimization features numerous points at which the binarization of the device is forced to increase by passing the current permittivity through a sigmoidal function. At iteration 512 the optimization is converted to a level-set optimization, which recovers the performance lost from that transition. Furthermore, the level-set optimization improves the performance of the device to approximately the same point as when the density-based optimization was 90\% binary, which is the approximate binarization when modelled by a level-set curve due to the continuous smoothing of the permittivity at simulation voxels that are intersected by the level-set curve.}
\label{fig:ch3-5:sugarcube_convergence}
\end{figure}

The convergence plots showing the average device figure of merit (FoM) and device binarization are shown in Fig. \ref{fig:ch3-5:sugarcube_convergence}. At various points the binarization is forced to increase by passing the permittivity through a sigmoidal function and changing the device permittivity to the output of this function. The FoM is then allowed to recover before repeating this discrete push in binarization. The device loses performance as the device becomes more binary. At iteration 512 the device is fully binarized and the optimization switches to a level-set optimization, which recovers some of the performance that was lost during the final phases of the density optimization. During the level-set optimization, the binarization is recorded as approximately 90\% because the boundaries of the device are smoothed out yielding a continuous permittivity value. This can intuitively be thought of as the level-set function passing through a simulation mesh voxel, which is modelled in the FDTD simulation as a dielectric volume average of the two materials. Therefore the density optimization is overly restrictive at binarization values above 90\%, since it does not allow for this type of border smoothing, instead modelling the device as a discretized grid of binary voxels. This explains why the level-set function is able to recover substantially more performance than what is lost at the initial conversion from density-based optimization to a level-set optimization, recovering to approximately the same average FoM value as the density-based optimization when it was 90\% binary.

\section{Device layers}

\begin{figure}[H]
\centering\includegraphics[width=15cm]{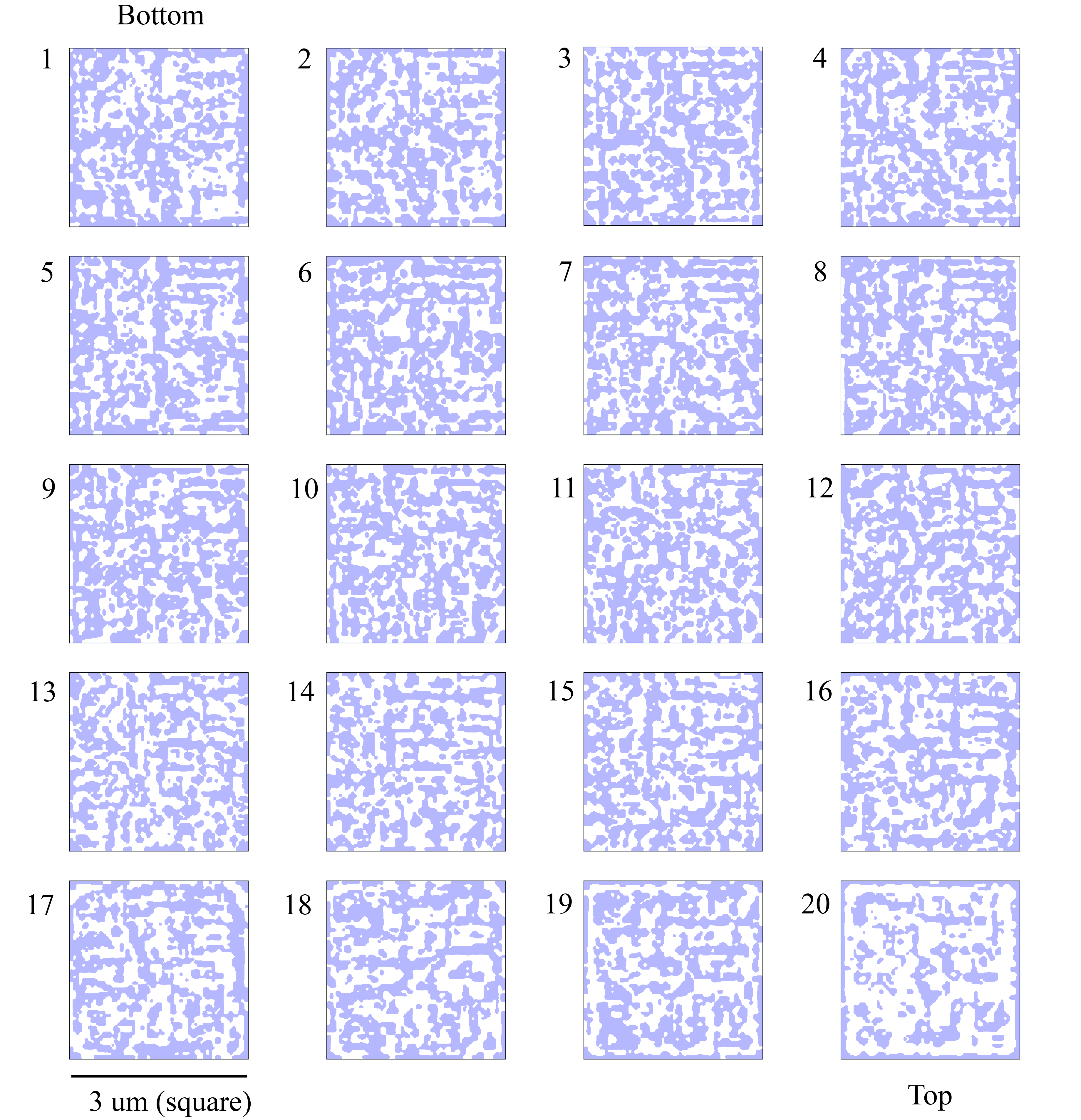}
\caption{\textbf{Device layers.} The shaded region is $TiO_2$, and the white is $SiO_2$. Layer 1 is the bottom layer nearest the sensor array. Each layer is 3x3$\mu m$.}
\label{fig:device_layers}
\end{figure}

\pagebreak

\section{Illustration of design procedure}

\begin{figure}[htbp]
\centering\includegraphics[width=12cm]{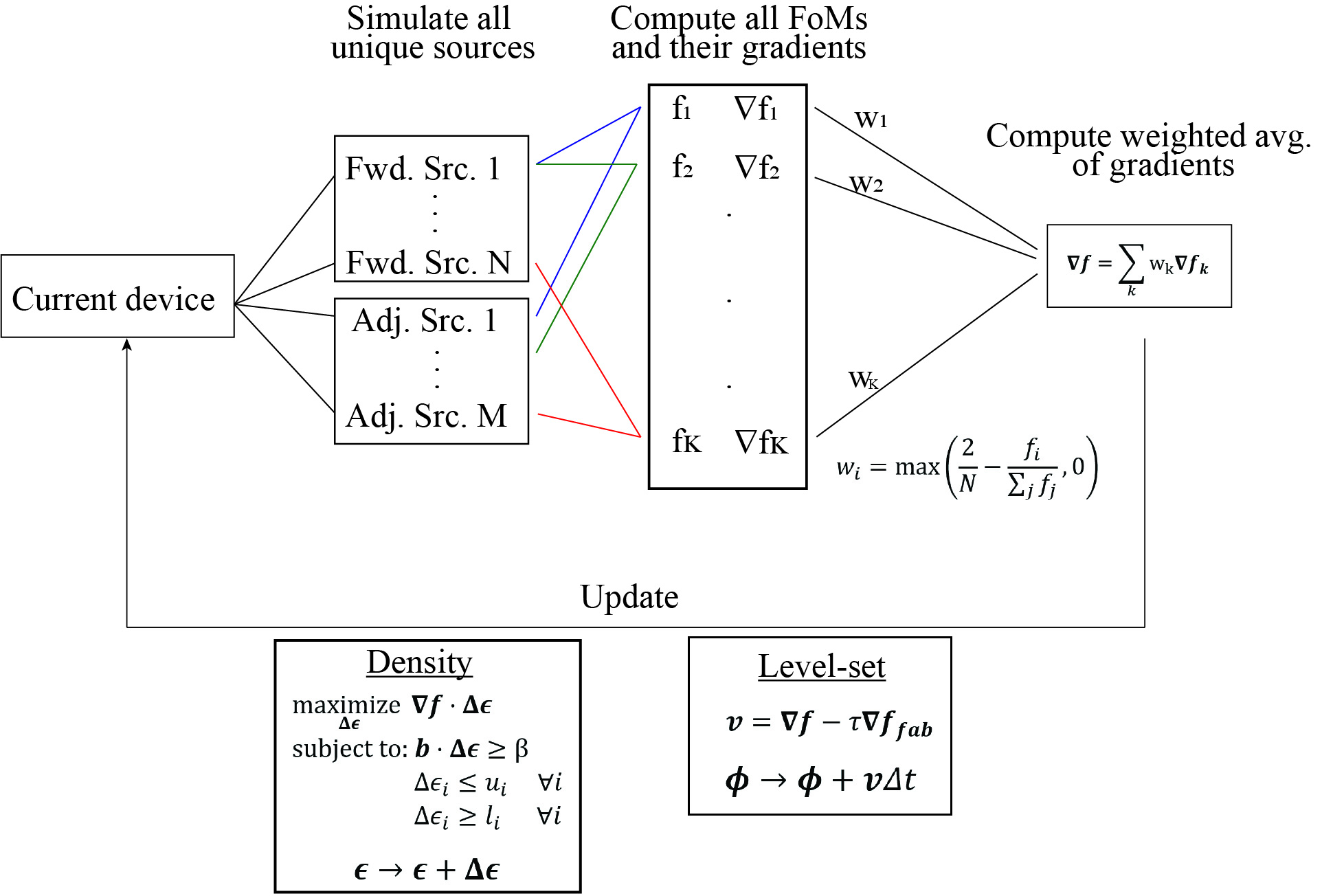}
\caption{An overview of the design methodology described in the main manuscript. For each iteration the process combines the results of multiple FDTD simulations, denoted forward and adjoint sources, to compute the gradient of many different FoMs. The various gradients are combined and used to update the structure using either a density-based method \cite{Ballew2021} or a level-set method \cite{Vercruysse2019}.}.
\label{Fig2}
\end{figure}

\section{Supplementary videos}

The videos provided as individual files show the electric field intensity at the focal plane of the device presented in the main manuscript. Each video shows how the output electric field power changes as the polarization, wavelength, and incident angle of the input fields are altered.

\textbf{Supplemental Video 1:} The polarization is swept continuously from x-polarization to y-polarization for 532 nm normally incident light.

\textbf{Supplemental Video 2:} The wavelength is swept continuously from 532 nm to 620 nm for y-polarized, normally incident light.

\textbf{Supplemental Video 3:} The incident angle $\theta$ is swept continuously from $-7^\circ$ to $+7^\circ$ for xz-polarized, 532 nm light with an azimuth angle of $\phi=0^\circ$.

\textbf{Supplemental Video 4:} The incident angle $\theta$ is swept continuously from $-7^\circ$ to $+7^\circ$ for xz-polarized, 532 nm light with an azimuth angle of $\phi=90^\circ$.

\bibliography{references}